\documentclass[twocolumn,aps,amsmath,amssymb,floatfix,superscriptaddress]{revtex4}
\usepackage{graphicx}
\usepackage{eucal}

\pdfoutput=1 

\begin{document}
\title{Exact solution of a two-type branching process: Models of tumor progression}

\author{Tibor Antal}
\address{School of Mathematics, University of Edinburgh, Edinburgh, EH9 3JZ, UK}
\author{P. L. Krapivsky}
\address{Department of Physics, Boston University, Boston, MA 02215, USA}

\begin{abstract}
An explicit solution for a general two-type birth-death branching process with one way mutation is presented. This continuous time process mimics the evolution of resistance to treatment, or the onset of an extra driver mutation during tumor progression. We obtain the exact generating function of the process at arbitrary times, and derive various large time scaling limits. In the simultaneous small mutation rate and large time scaling limit, the distribution of the mutant cells develops some atypical properties, including a power law tail and diverging average.
\end{abstract}


\maketitle

\section{Introduction}

Mathematical modeling has a long history in cancer research \cite{Frank:2007cr,Attolini:2009fk}. 
Stochastic models helped to establish the concept of multiple mutations in tumor progression \cite{Armitage:2004kx}, and led  to the understanding of tumor suppressor genes \cite{KNUDSON:1971uq,Kern:2002kx}. These models described the transitions between various stages of cancer \cite{Weinberg:2007ve,Vogelstein:2002bh} in an effective way, without considering population genetics. A more ``microscopic" approach is to model the stochastic evolution of a population of individual cells \cite{Nowak:2006oq}. The simplest such models are branching processes \cite{Harris:1963ij,Athreya:1972bh}, with countless biological applications \cite{Kimmel:2002qf,Haccou:2005dq} ranging from bacterial evolution \cite{Bartlett:1955ys,Zheng:1999vn} to cell homeostasis \cite{JCP:JCP1041250304}. Branching processes have also been used to model several aspects of tumor progression \cite{KENDALL:1960vn,Luebeck:2002qf,Iwasa:2006tg,Komarova:2006nx,Bozic:2010kx,Durrett:2010hc,Attolini:2009fk,Kimmel:2002qf}.

Can, however, simple branching processes provide a quantitative description of such complex biological phenomena as homeostasis -- the maintenance of healthy tissues?
A promising positive answer appeared in \cite{Clayton:2007ly}, where inducible genetic labeling was used to analyze the stochastic fate of the progenies of a single initial marked cell in the basal layer of epidermis in mice. The probabilities of finding clones of any given sizes at various times were fit remarkably well by a simple constant rate two-type branching process of progenitor ($A$) and post-mitotic cells ($C$).
The scheme of the process is as follows
\begin{equation}
\label{clayton}
 \begin{tabular}{ll}
 $A \to  AA$   ~~~~ &     rate $r$\\
 $A \to  AC$  &     rate $1-2r$\\
 $A \to  CC $              &     rate $r$\\
 $C \to  \emptyset$  &     rate $\gamma$\\
 \end{tabular}
\end{equation}

This model seemed unsolvable \cite{Clayton:2007ly} in the sense that there was no known explicit formula for the probability $P_{m,n}(t)$ of finding $m$ number of $A$ cells and $n$ number of $C$ cells for any finite time. 
We showed in \cite{Antal:2010fk} how to obtain an exact expression for the generating function of $P_{m,n}(t)$. Once the generating function is known, the probability $P_{m,n}(t)$ can be efficiently obtained numerically by fast Fourier transformation. 

The simplicity of the above model lies in the trivial behavior of $C$ cells -- which can only die but not proliferate -- and in the criticality of the process, that is the symmetry of birth and death rates of $A$ cells. Is it possible to solve more general two type birth-death processes explicitly where the second cell type can both die and proliferate? In this paper we present an explicit solution of such a general continuous time birth-death process, proposed by Kendall \cite{KENDALL:1960vn} to model the onset of mutations during tumor progression. In this model, $A$ cells can mutate irreversibly into $B$ cells according to 
\begin{equation}
\label{modeldef}
 \begin{tabular}{ll}
 $A \to  A A$   ~~~~ &     rate $\alpha_1$\\
 $A \to  \emptyset$  &     rate $\beta_1$\\
 $A \to  B $              &     rate $\nu$\\
 $B \to  BB$             &     rate $\alpha_2$\\
 $B \to  \emptyset$  &     rate $\beta_2$\\
 \end{tabular}
\end{equation}
We calculate the generating function that encapsulates the probability distribution $P_{m,n}(t)$ for finding $m$ copies of $A$ and $n$ copies of $B$ cells at time $t$.
Without loss of generality, we set the division rate of $A$ cells to one, $\alpha_1\equiv 1$, since this can always be achieved by simply rescaling time. We shall use the following shorthand notations for the rate differences
\begin{equation}
\label{deltas}
 \lambda_1 = 1-\beta_1-\nu , \quad 
 \lambda_2=\alpha_2-\beta_2
\end{equation}
which can be considered as the fitness values of the corresponding cell types.
Note that there is no restriction on the signs of parameters $\lambda_1$ and $\lambda_2$.

There are several possible applications of model \eqref{modeldef}. It can model the onset of a new mutation in an evolving population of progenitor cells (e.g.\ the $A$ cells of process \eqref{clayton}) of a healthy tissue. The new mutation can be either a neutral {\it passenger} ($\lambda_2=0$), or an advantageous {\it driver} ($\lambda_2>0$) \cite{Klein:2007ys, Klein:2010zr}. Model \eqref{modeldef} can also describe the onset of a chemotherapy resistant mutation in cancer \cite{Iwasa:2006tg,Komarova:2006nx}, corresponding to the case $\lambda_1=\lambda_2>0$, or provide a minimal model of metastasis formation, where $B$ cells play the role of metastasized cells \cite{Michor:2006zr,Dingli:2007ly,Yachida:2010uq}. When an advantageous (driver) mutation appears in a supercritical process ($\lambda_2>\lambda_1>0$), model \eqref{modeldef} provides a detailed description of the progression of a tumor towards malignancy. This scenario occurs in models that follow the onsets of multiple mutations \cite{Armitage:2004kx,Fearon:1990uq,Beerenwinkel:2007vn,Durrett:2010hc,Durrett:2010bs}. Such a multistage model was recently found to fit clinical data (on the number of driver versus passenger mutations in cancer) remarkably well \cite{Bozic:2010kx}. This demonstrates that branching processes can provide a quantitative descriptions of certain aspects of cancer.

Note that in model \eqref{modeldef} mutations happen at arbitrary times, which can be either a reasonable simplifying assumption, or a realistic feature when, for instance, mutations are caused by UV radiation \cite{Klein:2010zr}. More often, however, mutations happen during cell divisions \cite{Vogelstein:2002bh}, which case is described by the scheme
\begin{equation}
\label{mutatdiv}
 A \to  AB             \mbox{ ~~~~rate~} \nu
\end{equation}
instead of the third process of \eqref{modeldef}. The solution of this alternative process goes completely analogously to that of \eqref{modeldef}, and is discussed in Appendix \ref{divmut}.

The two-type process \eqref{modeldef} and other multi-type branching processes have been extensively studied \cite{Athreya:1972bh,Harris:1963ij,Kimmel:2002qf}, but the main focus was on the large time limit behavior. Our results, however, are exact for any finite time, which has relevance when comparing them to experiments. Multi-type branching process were  generally considered intractable as even in the simplest cases one arrives to (generally unsolvable) Riccati equations. Some notable exceptions are the two-type Luria-Delbruck models, which describe bacteria growth in a dish, where cell death can be neglected. Indeed, in the case of no death $\beta_1=\beta_2=0$, model \eqref{modeldef} with \eqref{mutatdiv} can be turned into a much easier Bernoulli equation, which leads to a simple solution \cite{Kimmel:2002qf}. Recently, however, we solved \cite{Antal:2010fk} a two-type branching process with cell death \eqref{clayton}, and this renewed the hope that Kendall's model of tumor growth is also solvable. We shall show that the processes \eqref{modeldef} and \eqref{mutatdiv} are also analytically tractable.

The rest of the paper is organized as follows. In Sec.~\ref{analysis} we obtain the explicit solution of the process \eqref{modeldef} in terms of generating functions. Special cases of the solution when either $A$ or $B$ cells behave critically are discussed in Sec.~\ref{critics}. Knowing the explicit solution allows us to calculate $P_{m,n}(t)$ for arbitrary times, but we also derive various asymptotic limits of the solution in Sec.~\ref{asymp}.
Finally, we draw conclusions in Sec.~\ref{conc}, and relegate some details to the Appendixes.

\section{General case}
\label{analysis}

Let us first try to develop an intuition for the system \eqref{modeldef} based on exact results regarding average densities. The average number of $A$ cells obeys the rate equation 
\begin{equation}
\label{m-av}
\frac{d \langle m\rangle}{d t} = \lambda_1\langle m\rangle
\end{equation}
Hence starting with one $A$ cell we get
\begin{equation}
\label{m-av-sol}
\langle m\rangle = e^{\lambda_1 t}
\end{equation}
Similarly the average number of $B$ cells evolves according to the rate equation 
\begin{equation}
\label{n-av}
\frac{d \langle n\rangle}{d t} = \nu\langle m\rangle + \lambda_2 \langle n\rangle
\end{equation}
Using \eqref{m-av-sol} and $\langle n\rangle(0)=0$ we solve \eqref{n-av} to yield
\begin{equation}
\label{n-av-sol}
\langle n\rangle = \nu\,\frac{e^{\lambda_2 t}-e^{\lambda_1 t}}{\lambda_2-\lambda_1}
\end{equation}
Higher order densities satisfy more complicated equations, but their solution is just a combination of exponentials.

The full description of the dynamics of the two-type branching process \eqref{modeldef} is provided by the  probabilities $P_{m,n}(t)$ of having $m$ copies of $A$ and $n$ copies of $B$ at time $t$. These probabilities satisfy the forward and backward Kolmogorov equations \cite{Athreya:1972bh}. In terms of the generating function 
\begin{equation}
\label{Pxy}
\mathcal{P}(x,y,t) = \sum_{m,n\ge 0} x^m y^n P_{m,n}(t)
\end{equation}
the {\it forward} equations reduce to a single linear partial differential equation. In this article, we shall use
{\it backward} Kolmogorov equations which in the present case are a pair of coupled non-linear ordinary differential equations ~\cite{Athreya:1972bh,Kimmel:2002qf}
\begin{subequations}
\label{maineqs}
\begin{align}
  \label{PA}
 \partial_t \mathcal{A} &= \mathcal{A}^2+\beta_1+\nu\mathcal{B}-(1+\beta_1+\nu)\mathcal{A}\\
  \label{PB}
 \partial_t \mathcal{B} &= \alpha_2 \mathcal{B}^2 + \beta_2 - (\alpha_2+\beta_2)\mathcal{B}
\end{align}
\end{subequations}
Here $\mathcal{A}$ and $\mathcal{B}$ are the realizations of the the generating function \eqref{Pxy} which only differ by their initial condition: the notation refers to the type of the single initial cell $A$ or $B$, that is the initial conditions are respectively
\begin{subequations}
\begin{align}
  \label{in-PA}
  \mathcal{A}(x,y, t=0) = x\\
  \label{in-PB}
  \mathcal{B}(x,y, t=0) = y
\end{align}
\end{subequations}
Both the positive (gain) terms and the negative (loss) terms in \eqref{PA} and \eqref{PB} correspond to the individual processes in \eqref{modeldef} in a straightforward way, reflecting on what happens to the initial cell. For example, since the initial $A$ cell divides into two $A$ cells at rate one, we find an $\mathcal{A}^2$ term with coefficient one on the right hand side of \eqref{PA}. Note that the forward Kolmogorov equation is a first-order hyperbolic partial differential equation and hence one can analyze it using the method of characteristics. The equations for the characteristics are mathematically identical to the 
backward Kolmogorov equations (10a) and (10b), so the analytical framework (forward or backward Kolmogorov equations) is the matter of choice; the latter description is a little more convenient as it does not involve intermediate steps related to the characteristics.

Equation \eqref{PB} contains only $\mathcal{B}$. It is tractable and its solution, subject to \eqref{in-PB}, reads 
\begin{equation}
\label{y-sol}
  \mathcal{B} = 1-\frac{\lambda_2}{\alpha_2(1-z)}\, , \quad z = \left[ 1- \frac{\lambda_2}{\alpha_2(1-y)}\right]e^{-\lambda_2 t}
\end{equation}

Plugging \eqref{y-sol} into \eqref{PA} we find that $X\equiv 1-\mathcal{A}$ satisfies a first order non-linear differential equation
\begin{equation}
\label{X-eq}
\frac{dX}{dt} = -X^2 +\lambda_1 X + \frac{\nu\lambda_2}{\alpha_2(1-z)}
\end{equation}
the Riccati equation. A useful way to treat Riccati equations is based on the standard trick \cite{Bender:1978cr} that turns a Riccati equation into a linear Sturm-Liouville equation. In the present case, this trick suggests to write
\begin{equation}
X \equiv \frac{d}{dt} \log Z = \frac{1}{Z}\, \frac{dZ}{dt}
\end{equation}
One finds that $Z$ satisfies the following Sturm-Liouville equation
\begin{equation}
\label{Z-eq}
\frac{d^2 Z}{dt^2} = \lambda_1\,\frac{dZ}{dt} + \frac{\nu\lambda_2}{\alpha_2(1-z)}\, Z
\end{equation}
Let us consider $\lambda_2>0$ for an instant. In the $t\to \infty$ limit, we have $z\to 0$ and hence Eq.~\eqref{Z-eq} turns into a linear equation with constant coefficients which admits an exponential asymptotic solution 
\begin{equation}
\label{Z-asymp-sol}
Z\propto e^{-\omega t},\quad  \omega^2  +\lambda_1\omega - \frac{\nu\lambda_2}{\alpha_2} = 0
\end{equation}
Using \eqref{y-sol} we can re-write the above asymptotic solution as 
$Z\propto z^{\omega/\lambda_2}$ when $z\to 0$. This suggests to seek the solution of 
Eq.~\eqref{Z-eq}  in the form
\begin{equation}
\label{Z-Phi}
Z(t) \equiv z^{\omega/\lambda_2}\,\Phi(z)
\end{equation}
for arbitrary $\lambda_2$.
Plugging \eqref{Z-Phi} into \eqref{Z-eq} we obtain
\begin{equation}
\label{Phi-eq}
z(1-z)\Phi'' + \left(1+\frac{2\omega+\lambda_1}{\lambda_2 }\right)(1-z)\Phi'=
\frac{\nu}{\alpha_2\lambda_2}\Phi 
\end{equation}
where the prime denotes the derivative with respect to $z$. Equation \eqref{Phi-eq} admits a solution in terms of the hypergeometric function. Indeed, the canonical hypergeometric equation reads  \cite{Bender:1978cr}
\begin{equation}
\label{hyper}
z(1-z)\,\Phi'' + [c-(a+b+1)z]\Phi' - ab\Phi = 0
\end{equation}
and it has two linearly independent solutions
\begin{equation}
\label{hypsols}
 F(a,b; c; z), \quad
 z^{1-c}F(a-c+1,b-c+1; 2-c; z) 
\end{equation}

Equation \eqref{Phi-eq} coincides with \eqref{hyper} if parameters $a,b,c$ satisfy
\begin{equation}
\label{ab-eq}
a+b = \frac{2\omega+\lambda_1}{\lambda_2 }\,, \quad ab = \frac{\nu}{\alpha_2\lambda_2}\, ,\quad c=a+b+1
\end{equation}
Using \eqref{Z-asymp-sol} and \eqref{hypsols} we arrive at 
\begin{equation}
\label{Z-sol}
\Phi(z) = F(a,b; c; z) + C z^{1-c} F(-b,-a;2-c;z)
\end{equation}
with parameters 
\begin{equation}
\label{abc}
a = \frac{\omega}{\lambda_2 }\,, \quad b = \frac{\omega+\lambda_1}{\lambda_2 }\,, \quad
c = 1+\frac{2\omega+\lambda_1}{\lambda_2 }
\end{equation}

To complete the solution \eqref{Z-Phi}, \eqref{Z-sol}--\eqref{abc} we need to know $\omega$ and $C$. 
Recall that $\omega$ is found from \eqref{Z-asymp-sol}; the proper root is
\begin{equation}
\label{omega}
\omega = -\frac{\lambda_1}{2} 
+ \sqrt{\left(\frac{\lambda_1}{2}\right)^2+ \frac{\nu\lambda_2}{\alpha_2}}
\end{equation}
Recalling previous definitions we have
\begin{equation*}
\mathcal{A}=1-\frac{1}{Z}\,\frac{dZ}{dt}= 1+\omega + \frac{\lambda_2 z}{\Phi}\,\frac{d\Phi}{d z}
\end{equation*}
We now compute the derivative of the hypergeometric function using identity \eqref{hyperdiff} given in Appendix \ref{formulas}. The original generating function $\mathcal{A}$ becomes
\begin{equation}
\label{genPA}
 \mathcal{A} = 1 + \omega + \lambda_2 \Psi(z)
\end{equation}
where we use the shorthand notation
\begin{equation}
\label{Psi} 
 \Psi(z) = \frac{z^cF_3(z)+C(1-c) F_2(z)+Cz F_4(z)}{z^{c-1} F_1(z)+C F_2(z)} 
\end{equation}
with
\begin{equation}
\label{F4}
\begin{split}
 F_1(z) &= F(a,b; c; z)\\
 F_2(z) &= F(-a,-b;2-c;z)\\
 F_3(z) &= \frac{ab}{c} F(1+a,1+b; 1+c; z)\\
 F_4(z) &= \frac{ab}{2-c} F(1-a,1-b,3-c; z)
\end{split}
\end{equation}
From \eqref{in-PA} we determine the value of the parameter 
\begin{equation}
\label{gencon}
 C = z_0^c\, \frac{\kappa F_1(z_0)-F_3(z_0)} {(1-c-\kappa z_0)F_2(z_0) + z_0 F_4(z_0)}
\end{equation}
with
\begin{equation}
\label{z0}
 \kappa = \frac{x-1-\omega}{\lambda_2 z_0} ~,\quad z_0=1-\frac{\lambda_2}{\alpha_2(1-y)}
\end{equation}

Now that we know the complete generating function $\mathcal{A}(x,y,t)$, we immediately know e.g.\ the generating function of the distribution of the number of $B$ cells, $\mathcal{A}(1,y,t)$, and that of the total number of $A$ and $B$ cells, $\mathcal{A}(x,x,t)$. We also know the probability of having no $B$ cells, $\mathcal{A}(1,0,t)$, or having no cells at all, $\mathcal{A}(0,0,t)$. We can extract $P_{m,n}(t)$ from this solution by using Cauchy's integral formula
\begin{equation}
\label{Cauchy}
P_{m,n}(t) = \frac{1}{(2\pi i)^2}
\oint \frac{dx}{x^{m+1}}\oint \frac{dy}{y^{n+1}}\,\mathcal{A}(x,y,t)
\end{equation}
Numerically, this inverse transformation can be efficiently performed via the fast Fourier transform algorithm \cite{Antal:2010fk}.

The solution of the alternative process \eqref{mutatdiv}, where mutations happen at cell divisions, is given in Appendix \ref{divmut}.

\section{Critical cases}
\label{critics}

Here we consider the special cases of critical dynamics of $A$ cells ($\lambda_1=0$), that of $B$ cells ($\lambda_2=0$), or both. If only $A$ cells are critical, all formulas of the previous section can be simply taken in the $\lambda_1\to 0$ limit. If, however, $B$ cells are critical ($\lambda_2=0$), expressions like \eqref{abc} are ill defined. This $\lambda_2=0$ case for general $\lambda_1$ is a bit cumbersome, but also implausible, as it describes a super- of subcritical population producing a completely neutral mutant. Hence we only consider here the bi-critical case $\lambda_1=\lambda_2=0$, that is when the birth and death rates of both cell types are balanced 
$1-\nu=\beta_1$, $\alpha_2=\beta_2$.

The behavior of critical $B$ cells can be obtained from \eqref{y-sol} by taking the $\lambda_2\to 0$ limit 
\begin{equation}
\label{y-sol-crit}
  \mathcal{B} = \frac{y+\alpha_2 t (1-y)}{1+\alpha_2 t (1-y)}
\end{equation}
We substitute this back to \eqref{PA}, and for $X=1-\mathcal{A}$ we arrive at
\begin{equation*}
 \frac{dX(t)}{dt} = -X^2 + \frac{\nu(1-y)}{1+\alpha_2 t (1-y)}
\end{equation*}
or equivalently
\begin{equation}
\label{critX}
 \frac{dX(\tau)}{d\tau} = -\frac{2\alpha_2\tau}{\nu}X^2 + \frac{2\nu}{\alpha_2\tau}
\end{equation}
in terms of the new variable
\begin{equation}
\label{criteps}
 \tau = \sqrt{\frac{\nu}{\alpha_2}\left[ t+\frac{1}{\alpha_2(1-y)}\right]}
\end{equation}
Again, by writing $X(\tau)=\frac{\nu}{2\alpha_2 t} \frac{d}{d\tau} \log Z(\tau)$, we simplify \eqref{critX} to 
$Z''-Z'/\tau-4Z=0$, which is solved by the modified Bessel functions
\begin{equation}
\label{Zsol-doublecrit}
 Z = \tau \left[ I_1(2\tau) + C K_1(2\tau) \right]
\end{equation}
up to an arbitrary constant.
By using identities \eqref{bessel4} for the modified Bessel functions,
solution \eqref{Zsol-doublecrit} can be transformed back to the original generating function
\begin{equation}
\label{critPA}
 \mathcal{A}(x,y,t) = 1- \frac{\nu}{\alpha_2\tau}\, \frac{I_0(2\tau) - C K_0(2\tau)}{I_1(2\tau)+CK_1(2\tau)}
\end{equation}
The parameter $C$ is fixed by the initial condition \eqref{in-PA}
\begin{equation}
\label{CritConst}
C = \frac{I_0(2\tau_0)-\alpha_2 \nu^{-1}\tau_0(1-x) I_1(2\tau_0)}
{K_0(2\tau_0) + \alpha_2 \nu^{-1}\tau_0 (1-x)K_1(2\tau_0)}
\end{equation}
with 
\begin{equation}
\label{CritTau}
 \tau_0 = \frac{1}{\alpha_2} \sqrt\frac{\nu}{1-y}
\end{equation}
This bi-critical case solution is somewhat simpler than the general solution \eqref{genPA}, as it only contains low index modified Bessel functions.


\section{Asymptotic Behaviors}
\label{asymp}

Now that we have the exact solution \eqref{genPA}, let us derive some asymptotic scaling behaviors.
The first task is to determine the right (or interesting) scaling limit. In the case of studying e.g.\ only the distribution $P_m(t)$ of one type of cells (either $A$ or $B$), we first guess the leading order large time asymptotic of the typical number of surviving cells $\chi(t)$, and we study the $t\to\infty$, $m\to\infty$ scaling limit, with $\hat{m}=m/\chi(t)$ kept constant. 
We define the scaling limit for the distribution $P_m(t)$ of the number of cells as $\chi(t) P_m(t) \to p(\hat{m})$, and note that the generating function becomes a Laplace transform, if we also take the $x\to1$ limit, while keeping $\xi=\chi(t) \log\frac{1}{x}\sim \chi(t)(1-x)$ constant
\begin{equation}
\label{Laplim}
\begin{split}
\mathcal A(x,t) &= \sum_{m\ge0} x^m P_m(t) = \sum_{m\ge 0} e^{-\hat{m}\xi} P_m(t)\\
&\to  \int_0^\infty e^{-\hat{m}\xi} p(\hat{m}) d\hat{m}
\end{split}
\end{equation}

The scaled density $p(\hat{m})$ has both a singular part describing the extinction of cells, and a regular part describing the surviving cells. The distribution of cells conditioned on survival is by definition $P^*_m(t)= P_m(t)/S(t)$, and we define its scaling limit as $\chi(t) P^*_m(t) \to p^*(\hat{m})$. Now by treating the first term of the sum in \eqref{Laplim} separately,
up to first order we obtain
\begin{equation}
\label{form}
\mathcal A(x,t)\to 1-s(t) + s(t) \mathcal{A}^*(\xi)
\end{equation}
with the Laplace transform
\begin{equation}
\mathcal{A}^*(\xi)=\int_0^\infty e^{-\hat{m}\xi} p^*(\hat{m}) d\hat{m}
\end{equation}
where the survival probability is asymptotically $S(t)\sim s(t)$. 
Consequently, $p(\hat{m}) = [1-s(t)] \delta(\hat{m}) + s(t) p^*(\hat{m})$. 
Our scaling describes the fate of all surviving cells if $\mathcal A^*(\xi)$ is the Laplace transform of a valid probability density, i.e.\ if $\mathcal A^*(0)=1$. But to obtain the right scaling of the surviving cell distribution, we first need to understand the asymptotic behavior of the survival probability.

\subsection{Survival probability}
\label{ext}

From the exact expressions \eqref{critPA} or \eqref{genPA} for $\mathcal{A}(x,y,t)$, we know the survival probability of $A$ cells $S_A(t)=1-\mathcal{A}(0,1,t)$, that of $B$ cells $S_B(t)=1-\mathcal{A}(1,0,t)$, and that of any cell $S(t)\equiv1-\mathcal{A}(0,0,t)$ at time $t$. Let us now use these exact expressions to understand their large time asymptotic behavior.

We first consider the bi-critical case $\lambda_1=\lambda_2=0$. It is immediate that that the survival probability of $A$ cells $S_A(t)\sim 1/t$ as $t\to\infty$. What is the survival probability of any cells $S(t)=1-\mathcal{A}(0,0,t)$? As $t\to\infty$, also $\tau\sim\sqrt{\nu t/\alpha_2}\to\infty$, while $C$ from \eqref{CritConst} remains constant. Now using the large argument limits \eqref{besselasympt} of the Bessel functions in \eqref{critPA}, we find that terms containing $C$ are asymptotically negligible and
\begin{equation}
\label{critsurvlim}
S(t) \sim \sqrt\frac{\nu}{\alpha_2 t}
\end{equation}
In this bi-critical case, of course, all survival probabilities $S_A$, $S_B$ and $S$ tend to zero, that is the eventual extinction is certain. But since $S$ is asymptotically much larger than $S_A$, it means that surviving cells are typically $B$ cells. In other words $S_B(t)\sim S(t)$, which is true in general for any parameter values due to the finite mutation rate $\nu$. This strange behavior, that neutral mutants are more likely to survive than non mutants, can be generalized for multiple type processes, with successive mutations $A_1\to A_2\to A_3\dots$. It turns out that that the survival probability of $A_i$ is asymptotically $S_i\propto t^{-1/2^{i-1}}$ \cite{Antal2011aa}.

In the supercritical case the survival probability is positive, hence the leading order asymptotic is constant $S(t)\to s_\infty\equiv 1-\mathcal{A}(0,0,\infty)$.
Let us first assume that $B$ cells are supercritical $\lambda_2>0$. In this case $z(t=\infty)=0$, and since from \eqref{abc} and \eqref{omega}
\begin{equation*}
c-1=\sqrt{\left(\frac{\lambda_1}{\lambda_2}\right)^2+\frac{4\nu}{\alpha_2 \lambda_2}}>0
\end{equation*}
we find that equation \eqref{genPA} reduces to 
\begin{equation}
\label{PA_simple}
\mathcal{A}(0,0,\infty)=1+\omega+\lambda_2(1-c) = 1-s_\infty
\end{equation}
Using \eqref{abc}--\eqref{omega} we can re-write the probability to end up in the state without cells as 
\begin{equation}
\label{NO}
s_\infty =  \frac{\lambda_1}{2} 
+ \sqrt{\left(\frac{\lambda_1}{2}\right)^2 + \frac{\nu\lambda_2}{\alpha_2}} = \omega+\lambda_1
\end{equation}
For $\lambda_2\to 0$ we recover the classical single type result $s_\infty = 1-\beta_1-\nu$ for $\beta_1+\nu<1$, and $s_\infty =0$ otherwise. For $\lambda_2<0$ there is no effect of $B$ cells on the survival probability, hence the result is the same as for $\lambda_2\to 0$. Note again that in the leading order $S_B(t)\sim S(t)$ for $t\to\infty$ due to the finite mutation rate $\nu$. 
Note also that the asymptotic survival probability $s_\infty$ can be obtained without the explicit knowledge of $\mathcal{A}(x,y,t)$, by solving the equations for $\mathcal{A}(0,0,t\to\infty)$ derived from \eqref{maineqs}. The above calculation therefore provides a check of self-consistency.

\subsection{Critical case}
\label{critlimits}

Many interesting limits are worth to investigate in the simplest bi-critical case, $\lambda_1=\lambda_2=0$, that is when the birth and death rates of both cell types are balanced $\alpha_2=\beta_2$ and $\beta_1=1-\nu$. We are going to describe the large time scaling behavior of $P_{m,n}(t)$ in three naturally arising regions of the $(m,n)$ plane: (i) corresponds to extinction $m=n=0$; (ii) corresponds to the survival of only $B$ cells $m=0, n>0$; and finally (iii) corresponds to the survival of both types $m,n>0$. For all cases of this bi-critical process, the average number of cells behave as
\begin{equation}
\label{critaver}
 \langle m \rangle =1 ,~~~  \langle n \rangle = \nu t
\end{equation}
but these averages include extinct lineages as well. To find the correct scaling for each regions we need to know the corresponding typical behaviors.

We understand the behavior in region (i) since in \eqref{critsurvlim} we already obtained that all cells go extinct asymptotically with probability $1-\sqrt{\nu/\alpha_2 t}$. Hence that main weight of $P_{m,n}$ is asymptotically concentrated at the origin.

Before exploring the $(m,n)$ plane further, as a warmup, let us first consider the behavior of $A$ cells alone, which sheds some light on the behavior of region (iii). Type $A$ cells are not affected by $B$ cells, and behave as a simple critical process, with generating function
 \begin{equation}
 \label{laponlyA}
 \mathcal{A}(x,1,t) = \frac{t(1-x)+x}{t(1-x)+1}
\end{equation}
The average (and typical) number of surviving cells is asymptotically $\chi(t)=t$, hence following the above method, we need to take the $t,m\to\infty$ and $x\to0$ limits with $\hat m=m/t$ and $\xi=t(1-x)$ constant. In this limit we recover scaling \eqref{form} with $\mathcal A^*(\xi)=1/(1+\xi)$, with $S_A\sim 1/t$. By an inverse Laplace transform we obtain an exponential density $p^*(\hat{m})=e^{-\hat{m}}$ for the scaled number of the surviving cells.

Now lets look at $B$ cells as well.
As we have shown in Sec.~\ref{ext}, for large times if there are surviving cells, they are typically $B$ cells only, hence the next largest weight corresponds to region (ii), that is to $m=0, n>0$. 
Therefore, we are interested in the distribution $P_n(t)$ of $B$ cells, irrespective of $A$ cells. The generating function of $P_n(t)$ is simply $\mathcal{A}(1,y,t)$. Next we need to guess the typical number of surviving $B$ cells in the asymptotic limit, which we set to $\chi(t)= \alpha_2 t$. The reason is that before the extinction of $A$ cells a few $B$ cells are produced, and they behave as a single type critical branching process of rate $\alpha_2$, which we just discussed above. This guess will be justified by the scaling function we obtain, i.e.\ by $\mathcal{A}^*(0)=1$. Alternatively, we could have just assumed a general algebraic anzatz for $\chi(t)$, and establish its linear time dependence from the scaling.

\begin{figure}
\centering
\includegraphics[scale=0.7]{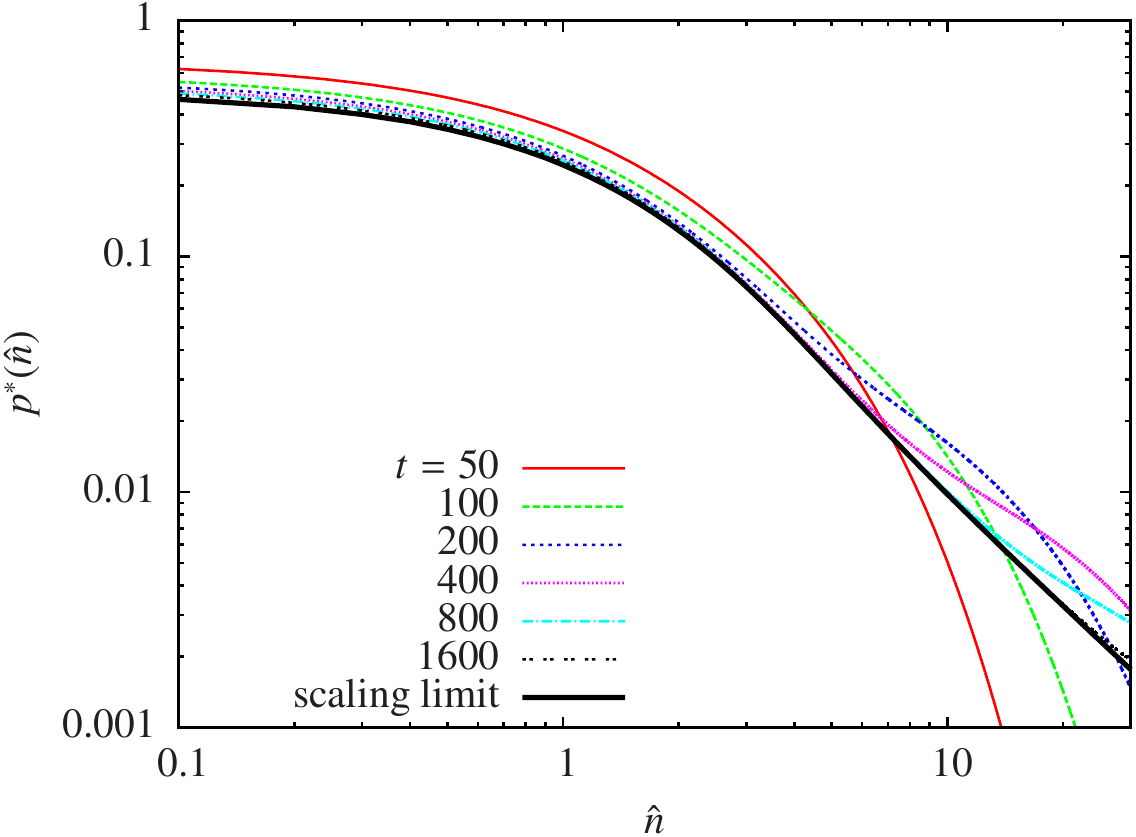}
\caption{Scaling of the probability $P_n(t)$ of having $n$ copies of $B$ cells at time $t$ in the bi-critical case, that is when death and birth rates are balanced for both cell types, $\alpha_1=\alpha_2=\beta_2=1$, while $\beta_1=1-\nu$. Here we present plots for mutation rate $\nu=0.1$ at several different times, calculated numerically from the exact generating function \eqref{critPA}. In the simultaneous limit $t,n\to\infty$ and $\hat n=n/\chi(t)$ being constant with $\chi(t)=\alpha_2 t$, the scaled distributions converge to a scaling limit $\chi(t)P_n(t)/S_B(t)\to p^*(\hat n)$ given by \eqref{critscalingfun}, where $S_B\sim\sqrt{\nu/\alpha_2 t}$.}
\label{crittimes}
\end{figure}

Hence we take the scaling limit $t\to\infty$, $y\to 1$, with $\eta=\alpha_2 t(1-y)$ kept constant, and from \eqref{criteps} and \eqref{CritTau} we find that both
\begin{equation}
\label{taulimits}
 \tau_0= \sqrt\frac{ \nu t}{\alpha_2\eta}, ~~~ \tau=\tau_0\sqrt{1+\eta}
\end{equation}
diverge as $t\to\infty$. Note that although $\tau_0$ depends only on $y$ in \eqref{CritTau}, we formally turned that into a $t$ dependence from $\eta=\alpha_2 t(1-y)$.
Therefore we use the large argument asymptotic of the modifies Bessel functions \eqref{besselasympt},
and note that $x=1$, to obtain from \eqref{CritConst} the asymptotic behavior of the parameter 
$C \sim e^{4\tau_0}/\pi$.
We see from \eqref{taulimits} that $\tau>\tau_0$ in the scaling limit, 
hence in \eqref{critPA} the terms proportional to $C$ become negligible, and the generating function simplifies to $A(1,y,t) = 1 - \nu/\alpha_2\tau$. From this we recover the scaling form \eqref{form} with
\begin{equation}
  \mathcal{A}^*(\eta) = 1-\sqrt{\frac{\eta}{1+\eta}}
\end{equation}
and with $S_B \sim \sqrt{\nu/\alpha_2 t}$. Note that $\mathcal{A}^*(0)=1$ as required. Now, by taking the inverse Laplace transform of the above expression (e.g. by integrating around the branch cut between $\eta=-1$ and 0), we find that the surviving $B$ cells are distributed as
\begin{equation}
\label{critscalingfun}
p^*(\hat n) = e^{-\hat n/2}  \left[I_0 (\hat n/2) - I_1(\hat n/2)\right]
\end{equation}
with the scaling variable $\hat n=n/\alpha_2 t$. Interestingly, the scaling function is independent of the mutation rate $\nu$. Note also that the above distribution has an algebraic tail $p^*(\hat n)\propto \hat n^{-3/2}$ for $\hat n\to\infty$, which implies an infinite average number of cells. This is understandable, since the average number of surviving $B$ cells grows as $\propto t^{3/2}$ from \eqref{critsurvlim} and \eqref{critaver}, which is faster than the typical value $\propto t$ we used for the scaling. The convergence to the scaling function $p^*(\hat n)$ for large times is illustrated in figure \ref{crittimes}. 

What is the distribution of cells when both type of cells are present, that is in the ``bulk" region (iii)? Note that the survival probability that cells of both types are still present at time $t$ scales asymptotically as $1/t$. Since $A$ cells behave critically, the number of surviving $A$ cells grows linearly with time, and consequently the typical number of $B$ cells grows as $\nu t^2$. This suggests to keep $m/t$ and $n/\nu t^2$ constant as $t\to\infty$. Hence in the generating function  \eqref{critPA} we should take the $t\to\infty$, $x,y\to 1$ limit, with $\xi=t(1-x)$ and $\eta=\nu t^2(1-y)$ constant. In this limit, from \eqref{criteps} and \eqref{CritTau}, we find again that both
\begin{equation}
 \tau_0\sim \frac{t\nu}{\alpha_2\sqrt{\eta}}, ~~~ 
 \tau = \tau_0 +\frac{\sqrt{\eta}}{2} + \mathcal{O}\left(\frac{1}{t}\right)
\end{equation}
diverge as $t\to\infty$.
Consequently, all terms of parameter $C$ of \eqref{CritConst} have the same leading order behavior, and we arrive at
\begin{equation}
 C\sim \frac{e^{4\tau_0}}{\pi} \frac{\sqrt\eta-\xi}{\sqrt\eta+\xi}
\end{equation}
Therefore all terms containing Bessel functions in \eqref{critPA} have the same order, and we finally find that
\begin{equation}
\label{critABlap}
 \mathcal{A}^*(\xi,\eta) = 1- \sqrt\eta\, \frac{\xi + \sqrt\eta\, \mathrm{th} \sqrt\eta}{\sqrt\eta + \xi\, \mathrm{th} \sqrt\eta}
\end{equation}
The scaling function of surviving $A$ and $B$ cells $p^*(m/t, n/\nu t^2)$
is the inverse Laplace transform of \eqref{critABlap} in both variables $\xi$ and $\eta$, which we could not obtain explicitely.

\subsection{Supercritical case}
\label{super}

There are many possible limits to take in a non-critial system, where $\lambda_1,\lambda_2\ne 0$. For example, one could consider all six possible orderings of $0, \lambda_1, \lambda_2$. Here we restrict our attention to the most interesting case, when an already advantegeous cell lineage produces an even fitter mutant, i.e.\ $\lambda_2>\lambda_1>0$. Again, we are interested in the distribution $P_n(t)$ of $B$ cells, irrespective of $A$ cells, which is given by $\mathcal{A}(1,y,t)$.

First, let us provide an intuitive reasoning for the scaling, which will be justified later.
Since $\lambda_2>\lambda_1$, the first surviving mutant produces the dominant portion of $B$ cells. A single $B$ cell eventually survives with probability $\lambda_2/\alpha_2$, and the average number of $B$ cells (including extinct lineages too) grows as $e^{\lambda_2 t}$. Consequently, the average number of surviving offspring of a single initial $B$ cell grows as $\chi(t)=\frac{\alpha_2}{\lambda_2} e^{\lambda_2 t}$ for $t\to\infty$, which we set as the typical number of surviving cells for the two type process. 

Hence, we study again $\mathcal{A}(1,y,t)$ in the scaling limit $t\to\infty$, $y\to 1$, with $\eta=\chi(t)\log 1/y\sim \chi(t)(1-y)$ being constant. Let us first determine the asymptotic value of the parameter $C$ given in \eqref{gencon}. In the scaling limit the parameter $z_0$ of \eqref{z0} diverges: $|z_0|=|1-e^{\lambda_2 t}/\eta| \to \infty$ for $\lambda_2>0$. Thus we need to determine the asymptotics of functions $F_j(z_0)$ which appear in \eqref{gencon}.  Since $\kappa z_0=-a$ (this follows from \eqref{z0} and \eqref{abc}; also recall that we consider $x=1$), the parameter $C$ is actually given by
\begin{equation}
\label{CF4}
 C = -z_0^{c-1}\, \frac{a F_1(z_0) + z_0 F_3(z_0)} {-b F_2(z_0) + z_0 F_4(z_0)}
\end{equation}
Note that $\lambda_1>0$ implies $a<b$ from \eqref{abc}. However, here we obtain the limit of $C$ for the case $a>b$, since this derivation is simpler, and the result is the same due to the analyticity of $C(a,b)$. Using the identities \eqref{hyperlimit} and \eqref{hyperspec} for the hypergeometric functions one finds 
\begin{equation}
\label{F4_asymp}
\begin{split}
 F_1(z_0) &\sim (-z_0)^{-b}\,\frac{\Gamma(c)\, \Gamma(a-b)}{\Gamma(1+a)\,\Gamma(a)} \\
 F_2(z_0) &\sim (-z_0)^a\,\frac{\Gamma(2-c)\, \Gamma(a-b)}{\Gamma(-b)\,\Gamma(1-b)} \\
 F_3(z_0) & \sim -\frac{b}{z_0}\,F_1(z_0)\\
 F_4(z_0) &  \sim  -\frac{a}{z_0}\,F_2(z_0)
\end{split}
\end{equation}
as $|z_0|\to \infty$. Substituting \eqref{F4_asymp} into \eqref{CF4} we obtain a neat expression for  the parameter
\begin{equation}
\label{Cab}
 C = (-1)^{a+b}\, \frac{b}{a}\,\frac{\Gamma(c)}{\Gamma(2-c)}\left[\frac{
 \Gamma(-b)}{\Gamma(a)}\right]^2
\end{equation}
Note that this result is valid for arbitrary values of $a$ and $b$.

It is clear from \eqref{genPA} that the generating function $\mathcal{A}(1,y,t)$ of $P_n(t)$ depends on $t$ and $y$ only through $C$ and $z$. We have just obtained $C$ in the scaling limit \eqref{Cab}, and from \eqref{y-sol} we find that $z\sim-1/\eta$. Hence analogously to the limiting form of \eqref{form}, from the definition \eqref{genPA} we obtain the Laplace transform of the surviving $B$ cell distribution in the scaling limit
\begin{equation}
\label{suplapgen}
 \mathcal{A}^*(\eta) = 1+ \frac{\omega+\lambda_2\Psi(-1/\eta)}{s_\infty}
\end{equation}
where $S(t)\to s_\infty=\omega+\lambda_1$ of \eqref{NO} is the leading order constant survival probability of $B$ cells, $\Psi(z)$ is given by \eqref{Psi} with $C$ of \eqref{Cab}. Therefore technically all we should do is to inverse the Laplace transform \eqref{suplapgen} to obtain the scaling function $p^*(\hat n)$ for the surviving $B$ cells, with 
\begin{equation}
\label{madhatter}
  \hat n= \frac{n}{\chi(t)} = n \frac{\lambda_2}{\alpha_2} e^{-\lambda_2 t} 
\end{equation}
Although this is easy to do numerically, analytically it is quite challenging, since the function $\Psi(z)$ involves the ratio of hypergeometric functions. This provides an example for a simple process where even the large time limit behavior is very complicated. 

One general comment concerns the behavior of the scaling function in the limit when the scaled variable $\hat n$ is large. We anticipate a generic exponential tail, 
$p^*(\hat n)\sim U\,e^{-u\hat n}$ as $\hat n\to\infty$, which implies that the Laplace transform has a simple pole at $\hat n=-u$ with residue $U$, that is,
\begin{equation}
\mathcal{A}^*(\eta) = \frac{U}{u+\eta}+\ldots 
\end{equation}
To determine $u, U$ we must therefore find a real and positive zero of function 
$z^{c-1}F_1(z)+CF_2(z)$. For special cases of integer $b$, we can explicitly compute the  scaling function $p^*(\hat n)$, and verify the exponential tail. We relegate these details to Appendix \ref{exam}.

\subsection{Small mutation, large time limit}

In many biological applications the mutation rate refers to a single base pair change, which is typically very small (in human DNA the probability of such mutation is around $5\times 10^{-10}$ per base-pair per cell division \cite{Jones:2008ve}). Hence in this subsection we study the joint large time small mutation rate limit. 

We consider again the case where both types are advantageous, and also $\lambda_2>\lambda_1>0$. We take the $\nu\to 0$ and $t\to\infty$ limits. As $\nu\to0$, we have $\lambda_1\to1-\beta_1$, and using 
\eqref{abc}--\eqref{omega} we obtain
\begin{equation}
\label{smallmutconsts}
a\sim a_1\nu, \quad a_1=\frac{1}{\alpha_2(1-\beta_1)}, \quad b\to b_0=\frac{1-\beta_1}{\lambda_2}
\end{equation}
where $0<b_0<1$. Also, we anticipate the corresponding scaled number of $B$ cells $\hat n\to0$, then $\eta\to\infty$, hence also $z\sim-1/\eta\to 0$. Therefore we take a simultaneous $\nu\to 0, \eta\to\infty$ limit, while keeping $\eta (a_1\nu)^{1/b_0}$ constant. The reason for keeping this particular combination constant will become clear later. 

The smallest order terms of \eqref{F4} are particularly simple due to \eqref{hyperseries}
\begin{equation*}
\begin{split}
 F_1(z) &\sim F_2(z) \sim 1\\
 F_3(z) &\sim\frac{a_1b_0}{1+b_0}\, \nu, \quad  F_4(z)\sim\frac{a_1b_0}{1-b_0}\, \nu
\end{split} 
\end{equation*}
while from \eqref{Cab}
\begin{equation*}
 C \sim  a_1\nu B_0 e^{\pi b_0}, \quad B_0=\frac{\pi}{\sin \pi b_0}
\end{equation*}
Substituting these limits into \eqref{Psi} we obtain
\begin{equation}
\label{psilowmut}
\Psi(-1/\eta) = \frac{-b_0}{1+B_0^{-1}\left[\eta (a_1\nu)^{1/b_0}\right]^{-b_0}}
\end{equation}
It is clear that we obtain the only nontrivial limit when the new scaling variable $\zeta= (a_1\nu)^{1/b_0} \eta=\chi(t,\nu)(1-y)$ is kept constant, with
\begin{equation}
\label{typtnu}
 \chi(t,\nu) = \frac{\alpha_2}{\lambda_2} e^{\lambda_2 t} (a_1\nu)^{1/b_0}
\end{equation}
Since for $\nu\to0$ the survival probability \eqref{NO} is $s_\infty\to1-\beta_1$, and $\omega\sim \lambda_2 a_1\nu$, the scaling function $\mathcal{A}^*$ from \eqref{suplapgen} becomes
\begin{equation}
\label{ssLap}
 \mathcal{A}^*(\zeta) = \frac{1}{1+B_0 \zeta^{b_0}}
\end{equation}
This scaling function depends only on the single parameter $0<b_0<1$. 
The corresponding scaled number of $B$ cells is $\tilde n = n/\chi(t,\nu)$, with the scaling function $p^*(\tilde{n})$.
This generating function has been studied numerically in \cite{Iwasa:2006tg}, and it has already been derived in a very appealing approximate model in \cite{Durrett:2010hc}.

\begin{figure}
\centering
\includegraphics[scale=0.7]{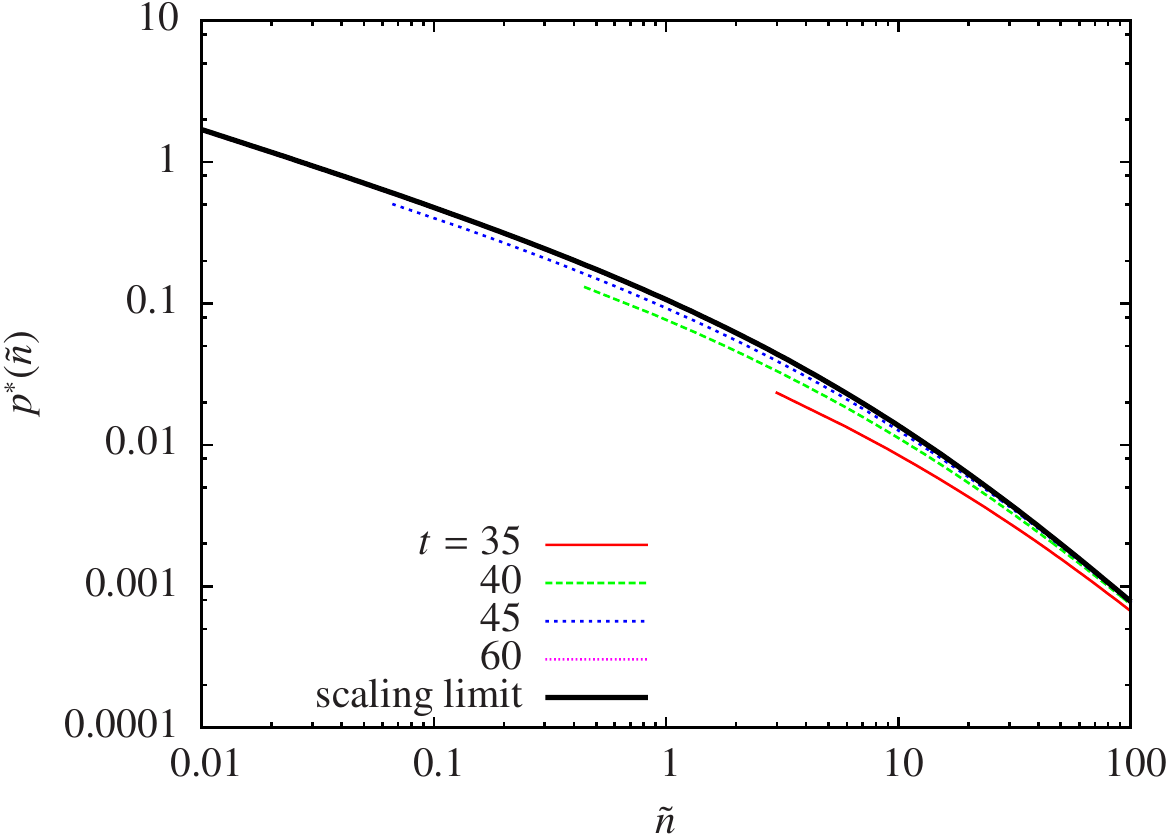}
\caption{Scaling of the probability $P_n(t)$ of having $n$ copies of $B$ cells at time $t$ in the supercritical case for the simultaneous limit of large time, large number of cells and small mutation rate, $t, n\to\infty, \nu\to 0$ with $\tilde n=n/\chi(t,\nu)$ constant and $\chi(t,\nu)$ given by \eqref{typtnu}. We chose $\alpha_1=1$,  $\beta_1=0.81$, $\alpha_2=1.29$, $\beta_2=0.91$, which correspond to the scaling parameter $b_0=1/2$. Plots are for mutation rate $\nu=0.0001$ at different times, calculated numerically from the exact generating function \eqref{genPA}. The scaled distributions converge to the limit function $\chi(t,\nu)P_n(t)/S_B(t)\to p^*(\tilde n)$ given by \eqref{smutexp}, where $S_B\to1-\beta_1$.}
\label{suptimes}
\end{figure}

Let us explore the properties of $p^*(\tilde{n})$. For the special case $b_0=1/2$ we can evaluate the inverse Laplace transform of $\mathcal{A}^*(\zeta)$ from \eqref{ssLap} explicitely
\begin{equation}
\label{smutexp}
 p^*(\tilde{n}) = \frac{1}{\pi^2} \left[ \sqrt{\frac{\pi}{\tilde n}} - e^{\tilde n /\pi^2} \mathrm{Erfc} \frac{\sqrt{\tilde n}}{\pi} \right]
\end{equation}
For general $b_0$, from the large $\eta$ limit of $\mathcal{A}^*(\zeta)$, we can obtain the small $\tilde n$ series of the scaling function
\begin{equation}
\label{smuts}
 p^*(\tilde n) = \sum_{j\ge 1}  \frac{ -1}{(-B_0)^j \Gamma(b_0j)}\,  \tilde{n}^{b_0j-1} 
\end{equation}
where $B_0>0$ is given by \eqref{psilowmut}. Indeed, by taking the Laplace transform of the formal series expansion of $p^*(\tilde n)$ around $\tilde n=0$ and equating it term by term to the series expansion of $\mathcal{A}^*(\zeta)$ around $\zeta=\infty$, we obtain \eqref{smuts}.
Conversely, from the small $\eta$ asymptotic of $\mathcal{A}^*(\zeta)$ we obtain the large $\tilde n$ series
\begin{equation}
\label{smutl}
 p^*(\tilde n) = \sum_{j\ge 1} \frac{(-B_0)^j}{ \Gamma(-b_0 j)}\,  \tilde{n}^{-b_0j-1}
\end{equation}
This is achieved similarly to the small $\tilde n$ series, but some care is needed. To avoid certain singularities, the series expansion of the derivative $\frac{d}{d\zeta}\mathcal{A}^*(\zeta)$ should be compared to the Laplace transform of the expanded $\tilde n p^*(\tilde n)$. Note that the small $\tilde n$ series \eqref{smuts} has an infinite radius of convergence. Conversely, the large $\tilde n$ series \eqref{smutl} has zero radius convergence, but it is asymptotic and hence truncating it after a finite number of terms provides an excellent approximation \cite{Bender:1978cr}.

\begin{figure}
\centering
\includegraphics[scale=0.7]{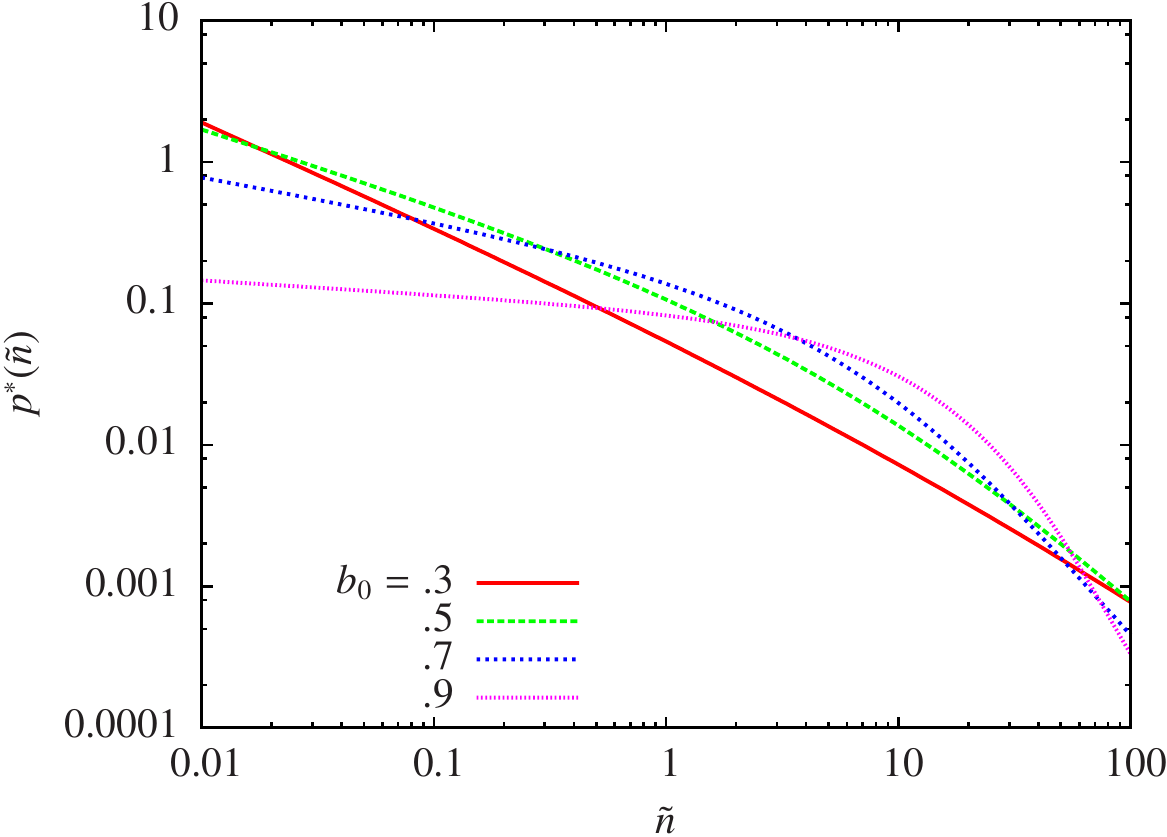}
\caption{Scaling function $p^*(\tilde n)$ of $B$ cells for the supercritical case in the large time small mutation limit evaluated from \eqref{smuts}. We plotted this function at four values of the single parameter $b_0$. Note the explicit formula \eqref{smutexp} for $b_0=1/2$. The small and large $\tilde n$ asymptotic power law behavior can be observed.}
\label{supscales}
\end{figure}

There are two surprising features of the scaling function $p^*(\tilde n)$. First, from the small $\tilde n$ series \eqref{smuts} we find that the density is singular at $\tilde n=0$ as $p^*(\tilde n)\propto \tilde{n}^{b_0-1}$, where $-1<b_0-1<0$. 
Second, from the large $\tilde n$ series \eqref{smutl} we find that the density has a power law tail $p^*(\tilde n)\propto \tilde{n}^{-(1+b_0)}$, with the exponent being $1<1+b_0<2$. This means that this density function has an infinite average. The scaling for large times is demonstrated in figure \ref{suptimes}, while the scaling function $p^*(\tilde n)$ is depicted in figure \ref{supscales} for several values of its only parameter $b_0$.

The power law tail of $p^*(\tilde{n})$ has been observed numerically in \cite{Iwasa:2006tg}. Note also that we have already encountered a limit density with infinite average in Sec.~\ref{critlimits}, for a critical process for arbitrary mutation rate. Here, however, a power law density appeared in a non-critical situation, where we expect exponential tails in general (see Appendix \ref{exam}).

\section{Conclusions}
\label{conc}

In this paper we presented an explicit solution for a general two-type birth-death process with one-way mutations. This process was proposed by Kendall \cite{KENDALL:1960vn} to model the onset of a driver mutation in an evolving cell population. We computed the generating function \eqref{genPA}--\eqref{F4} that encapsulates the probability distribution $P_{m,n}(t)$. The generating function can be turned into the exact probabilities $P_{m,n}(t)$ by a fast Fourier transform \cite{Antal:2010fk}.

Our explicit result \eqref{genPA}--\eqref{F4} is not easy to grasp, as it involves the ratio of hypergeometric functions. It becomes somewhat simpler \eqref{critPA} in the bi-critical case, that is when both cell types behave critically $\lambda_1=\lambda_2=0$. The complete solutions, however, contain all information, for instance one immediately knows the survival probability of cells, or any order moments of the cell distribution. We have also extracted several interesting limits from the exact solutions. In the bi-critical case we showed in Sec.~\ref{critlimits} how different types of scaling apply in different regions of the $(m,n)$ plane. For the onset of an advantegeous mutation, $\lambda_2>\lambda_1>0$, we derived the large time scaling limit, which limit function still involved hypergeometric functions. This demonstrates that the large time limit behavior can be still very complicated. For certain special cases though, we derived the explicit scaling functions, with generic exponential tails. Conversely, in the simultaneous large time and small mutation limit, the scaling function has a power law tail, with infinite average value. As a consequence, an enormous number of samples are needed in simulations to recover the exact (finite time) average values.

There are of course numerous interesting open questions of two-type branching processes which deserve further attention. The long term description of tumor development though must involve more than two cell types
corresponding to multiple stages of tumors \cite{Armitage:2004kx,Fearon:1987fk,Fearon:1990uq}. These multi-type branching processes are likely too cumbersome for explicit solutions, and one needs simplifying assumptions to deduce the relevant asymptotic behavior \cite{Durrett:2010hc,Antal2011aa}. Nevertheless, some effective models recently proved to be remarkably successful in connecting theory and clinical data \cite{Bozic:2010kx}.

\section*{Acknowledgments}

We thank Ivana Bozic for useful suggestions. PKL gratefully acknowledges financial support from NSF grant CCF-0829541.

\appendix

\section{Mutation at cell divisions}
\label{divmut}

Here we briefly discuss a version of the two type process \eqref{modeldef}, where mutations happen at cell divisions \eqref{mutatdiv}. In \eqref{PA} the term  $\nu \mathcal{A}$ should be replaced by  $\nu \mathcal{A} \mathcal{B}$; all the rest in \eqref{PA}-\eqref{PB} remains the same. We follow the exact same steps as for the original model, and we arrive at the same solution \eqref{genPA}-\eqref{z0}, but with parameters
\begin{equation*}
a = \frac{\omega}{\lambda_2 }\,, \quad b = \frac{\omega+1-\beta_1}{\lambda_2 }\,, \quad
c = 1+a+b-\frac{\nu}{\alpha_2}
\end{equation*}
and
\begin{equation*}
\omega = -\left( \frac{\lambda_1}{2} + \frac{\nu\beta_2}{2\alpha_2} \right) 
+ \sqrt{\left( \frac{\lambda_1}{2} + \frac{\nu\beta_2}{2\alpha_2}  \right)^2
+\frac{\nu\lambda_2}{\alpha_2}}
\end{equation*}
instead of \eqref{abc} and \eqref{omega}. Also, while $F_1(z)$ and $F_3(z)$ are unchanged from \eqref{F4}, we need $F_2(z)$ and $F_4(z)$ in their general forms 
\begin{equation*}
\begin{split}
 F_2(z) &= F(1+a-c,1+b-c;2-c;z)\\
 F_4(z) &= \frac{(1+a-c)(1+b-c)}{2-c} \\
 & \times F(2+a-c,2+b-c,3-c; z)
\end{split}
\end{equation*}
since in \eqref{F4} we simplified these general formulas by using $c= 1+a+b$, which is not true in the present case.

\section{Case of integer $b$}
\label{exam}

In Sec.~\ref{super} we conjectured that the scaled distribution $p^*(\hat n)$ has an exponential tail in general. In this Appendix we review a few concrete examples for the special case of integer $b$ values. Recall that we assumed $\lambda_2>\lambda_1>0$ in Section \ref{super}, which implies $b>a>0$ from \eqref{abc}. We do not impose any other restriction on $a$.

Equation \eqref{Cab} shows that $C$ diverges when $b$ is a non-negative integer. This case corresponds to particular values of the mutation rate
\begin{equation*}
 \nu = \frac{1-\beta_1-b\lambda_2}{1-(b\alpha_2)^{-1}}
\end{equation*}
from the definitions \eqref{Z-asymp-sol} and \eqref{abc}, provided that $0<\nu<1-\beta_1$.
Our sole reason to consider these special mutation rates is to make the problem more tractable, and we anticipate that the behavior is qualitatively similar for arbitrary mutation rates. Indeed, when $C=\infty$, equation \eqref{Psi} simplifies to
\begin{equation*}
\Psi= 1-c+z\,\frac{F_4(z)}{F_2(z)}
\end{equation*}
and hence  \eqref{suplapgen} becomes
\begin{equation*}
\mathcal{A}^*(-1/z) = 1+ \frac{1}{s_\infty} \left[ \omega + \lambda_2(1-c) + \lambda_2\,z\,\frac{F_4(z)}{F_2(z)}\right]
\end{equation*}
Note that $\eta=-1/z$ in this limit.
The constant term on the right-hand side vanishes, see \eqref{PA_simple}, and therefore we finally arrive at
\begin{equation}
\label{Lap}
\mathcal{A}^*(-1/z) = \frac{\lambda_2 z F_4(z)}{s_\infty F_2(z)}
= \frac{z}{b}\,\frac{d\log F_2}{dz}
\end{equation}
Here we also used that $s_\infty/\lambda_2=b$ from \eqref{abc} and \eqref{NO}. Note that this relation provides another interpretation for integer $b$. Let us consider now concrete examples. Since $b>0$, the simplest possible choice is $b=1$.

\subsubsection{$b=1$}

Since $c=a+b+1=a+2$ we conclude that
\begin{equation*}
F_2=F(-1,-a;-a;z) = 1-z, \quad F_4=-1
\end{equation*}
Therefore Eq.~\eqref{Lap} becomes $\mathcal{A}^*(-1/z) = z/(z-1)$. In other words, $\mathcal{A}^*(\eta) = 1/(1+\eta)$, from which $p^*(\hat n)=\exp(-\hat n)$. Thus in this case the scaled distribution of $B$ cells is a pure exponential.

\subsubsection{$b=2$}

We have $c=a+b+1=a+3$ and
\begin{eqnarray*}
F_2&=&F(-2,-a;-1-a;z)=1-\frac{2a}{a+1}\,z+\frac{a-1}{a+1}\,z^2\\
F_4&=&\frac{dF_2}{dz} = -\frac{2a}{a+1}+2\,\frac{a-1}{a+1}\,z
\end{eqnarray*}
Equation \eqref{Lap} becomes
\begin{equation*}
\mathcal{A}^*(\eta) = \frac{a-1+a\eta}{a-1+2a\eta+(a+1)\eta^2}
\end{equation*}
and its inverse Laplace transform 
\begin{equation*}
p^*(\hat n) = \frac{1}{2} \left[e^{-\hat n}+
\frac{a-1}{a+1}\,\,\exp\!\left(\!-\hat n\, \frac{a-1}{a+1}\right)\right]
\end{equation*}
is a combination of two exponents. It is possible to obtain explicit expressions for larger $b$ values in particular, but let us consider the problem now in general.

\subsubsection{Arbitrary integer $b$}

In this case $F_2=F(-b,-a;1-b-a;z)$ is a polynomial of the order $b$; hence $F_4$ is a polynomial of order $b-1$. It is easy to see that $\mathcal{A}^*(\eta)$ is the ratio of a polynomial of $\eta$ of the order $b-1$ to the polynomial of $\eta$ of the order $b$. Therefore one can write an {\em exact} expansion
\begin{equation*}
\mathcal{A}^*(\eta)=\sum_{j=1}^b \frac{U_j}{u_j+\eta}
\end{equation*}
Thus when $b$ is integer, the scaling function is a combination of $b$ exponentials:
\begin{equation*}
p^*(\hat n)=\sum_{j=1}^b U_j\,e^{-u_j\hat n}
\end{equation*}

\section{Useful formulas}
\label{formulas}

Here we list a few identities and limits of special functions from \cite{Gradshtein:2000nx,Olver:2010oq}.
The modified Bessel functions satisfy
\begin{equation}
\label{bessel4}
 \begin{split}
 2 I_1'(z) &= I_0(z)+I_2(z)\\
 -2 K_1'(z) &= K_0(z)+ K_2(z)\\
 2I_1(z)/z &= I_0(z)-I_2(z)\\
 - 2K_1(z)/z &= K_0(z) - K_2(z)
 \end{split}
\end{equation}
and their $z\to\infty$ asymptotic behavior is
\begin{equation}
\label{besselasympt}
 I_a(z)\sim \frac{1}{\sqrt{2\pi z}} e^z, ~~~ K_a(z)\sim \sqrt\frac{\pi}{2z} e^{-z}
\end{equation}

The hypergeometric function has the series expansion
\begin{equation}
\label{hyperseries}
 F(a,b; c; z) = 1 + \frac{ab}{c} \frac{z}{1!} + \frac{a(a+1)b(b+1)}{c(c+1)} \frac{z^2}{2!} +\dots
\end{equation}
and the derivative
\begin{equation}
\label{hyperdiff}
 \frac{d}{dz} F(a,b; c; z) = \frac{ab}{c} F(1+a,1+b; 1+c; z) 
\end{equation}
We use the transformation formulas
\begin{equation}
\label{hyperlimit}
\begin{split}
 F(a,b;c;z) &= (1-z)^{-a} F(a,c-b;c;1-1/z)\\
 &= (1-z)^{-b} F(b,c-a;c;1-1/z)
\end{split}
\end{equation}
and the identity
\begin{equation}
\label{hyperspec}
 F(a,b;c;1) = \frac{\Gamma(c)\Gamma(c-a-b)}{\Gamma(c-a)\Gamma(c-b)}
\end{equation}
for $\mathrm{Re}\, c > \mathrm{Re}(a+b)$.

\end{document}